\documentclass[acmsmall]{acmart}

\usepackage{amsmath,amssymb,amsfonts}
\usepackage{algorithmic}
\usepackage{graphicx}
\usepackage{textcomp}
\usepackage{hyperref}
\usepackage{tikz}
\usepackage{xcolor}
\usepackage{tabularx}
\usepackage{multicol}
\usepackage{multirow}
\usepackage{booktabs} 
\usepackage{color}
\usepackage[utf8]{inputenc}

\newcommand{\fakeparagraph}[1]{\smallskip\noindent\textbf{#1.}}
\newcommand*\circled[1]{\tikz[baseline=(char.base)]{
            \node[shape=circle,draw,inner sep=1pt] (char) {#1};}}

\AtBeginDocument{%
  \providecommand\BibTeX{{%
    \normalfont B\kern-0.5em{\scshape i\kern-0.25em b}\kern-0.8em\TeX}}}

\begin{document}




\title{Demonstration of a Cloud-based Software Framework for Video Analytics Application using Low-Cost IoT Devices}


\author{Bhavin Joshi}
\affiliation{\institution{Redspark Technologies, India}}
\email{bhavin@redsparkinfo.co.in}

\author{Tapan Pathak, Vatsal Patel, Sarth Kanani}
\email{tapan.pce18@sot.pdpu.ac.in,vatsal.pce18@sot.pdpu.ac.in, sarth.kce18@sot.pdpu.ac.in, shailesh.ace16@sot.pdpu.ac.in}
\affiliation{%
  \institution{Pandit Deendayal Petroleum University}
  \city{Gandhinagar}
  \country{India}
}

\author{Pankesh Patel, Muhammad Intizar Ali, John Breslin}
\email{pankesh.patel@insight-centre.org}
\email{ali.intizar@nuigalway.ie}
\email{john.breslin@nuigalway.ie}
\affiliation{%
  \institution{Confirm SFI Research Centre for Smart Manufacturing, Data Science Institute, NUI Galway, Ireland}
}

\begin{abstract}
The design of products and services such as a Smart doorbell, demonstrating video analytics software/algorithm functionality, is expected to address a new kind of requirements such as designing a scalable solution while considering the trade-off between cost and accuracy; a flexible architecture to deploy new AI-based models or update existing models, as user requirements evolve; as well as seamlessly integrating different kinds of user interfaces and devices. To address these challenges, we propose a smart doorbell that orchestrates video analytics across Edge and Cloud resources. The proposal uses AWS as a base platform for implementation and leverages \textbf{C}ommercially \textbf{A}vailable \textbf{O}ff-\textbf{T}he-\textbf{S}helf~(COTS) affordable devices such as Raspberry Pi in the form of an Edge device. 
\end{abstract}





\maketitle

\section{Introduction}
Doorbells have been playing an important role in protecting the security of modern homes since they were invented. Existing smart doorbells~\cite{Delaney2020} often have limited processing capabilities due to either their portability requirements or to lower the cost of installation, deployment and maintenance. On the other hand, the rise of DL/CV algorithms~(to improve the object-detection accuracy at the Edge) demands powerful but expensive devices~(e.g., NVIDIA Jetson Nano, Google Coral USB Accelerator). To address this trade-off, we implement a smart doorbell that can orchestrate the placement of different components while keeping the trade-off between cost and detection accuracy in mind~\cite{iot-conf-2020}\footnote{This demo paper is backed by a conference paper, titled ``\textit{A Distributed Framework to Orchestrate Video Analytics Across Edge and Cloud: A Use Case of Smart Doorbell}'', at the 10th International Conference on the Internet of Things~(IoT), 2020.}.


The design of products and services such as Smart doorbell are expected to address a new kind of requirements /challenges, such as:

\begin{itemize}
    \item  Designing a scalable solution that can orchestrate a placement of different components, considering the resource constraints factors and addressing the high-accuracy in object detect requirements.

    \item A flexible architecture to deploy new AI-based models or update existing models, as user requirements evolve.

    \item Detecting one or multiple objects in real-time, anywhere and anytime, as the prime objective of the doorbell is to make the home secured.

    \item Seamlessly integrating with different kinds of devices and user interfaces (e.g. Amazon Alexa, Google Assistant, etc.), different cameras, smart phones.
\end{itemize}

Existing smart doorbells are bespoke and proprietary, which offer limited reusability and composability~\cite{scalable-framework, chauhan2016development, GYRARD2017305, 10.1145/3041021.3054736, 7460669, intizar:emse-01644333}.  It would increase both  the time and effort to integrate Deep Learning-based computer vision approaches for IoT devices.  For instance,  the users may would like to integrate new DL-based models to prevent package theft ~\cite{CNBC2020} or he may want to develop a collaborative learning among smart doorbell devices to detect suspicious activities around his home. To address such evolving requirements, we would need a transparent design that can offer the flexibility to deploy new features with a minimal efforts. In addition to this, the transparent design can greatly aid the use of this technology in multiple application domains. 


By demoing\footnote{An online video demonstration is available at this URL: \texttt{\url{https://www.youtube.com/watch?v=42mx4Z2PDwA}}.} the Smart doorbell and letting attendees interact with it, our objective is to start a discussion with experts in multiple relevant disciplines about the potential uses of this technology outside the realm of smart home and smart surveillance. We believe that the IoT conference is the perfect venue to open up this conversation.

\section{System Overview and Implementation}
The proposed architecture consists of three layers based on their functions: \textit{Device Layer}, \textit{Edge Layer}, and \textit{Cloud Layer}. 

\fakeparagraph{Device Layer} Each device interfaces with a camera module to capture images and has a PIR sensor to detect the motion of an object. The integration of motion sensors with the device allows to process data only when there is a motion. Each device implements AWS IoT-based device registration and X.509 certificate-based authentication functionality that allows the user to interact with the device anywhere and anytime.

\begin{figure}[h]
  \centering
  \includegraphics[width=\linewidth]{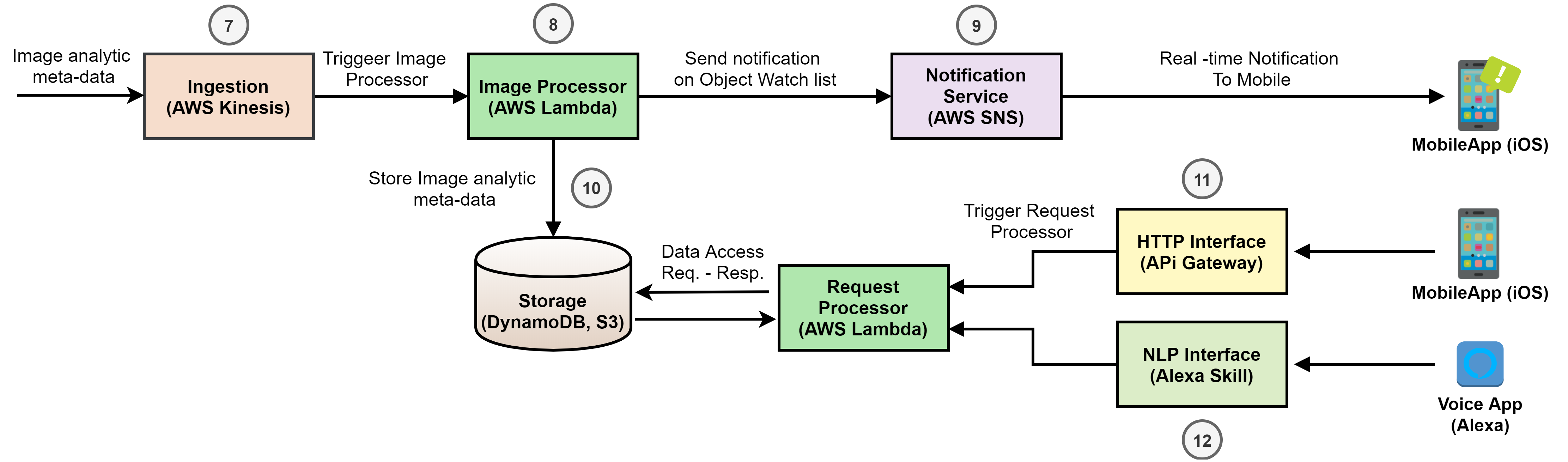}
  \caption{Logic Flow of Cloud Layer. It shows the components involved in the architecture and the technology (mentioned in round bracket) used to implement the component.}
  \label{fig:cloud}
\end{figure}

\fakeparagraph{Edge Layer} The frame sampling component samples a frame off of a live video stream from the attached camera and then packages and sends raw footage to the object detection component, which implements AWS Rekognition\footnote{\url{https://aws.amazon.com/rekognition/}} APIs for object recognition and detection.  The image analysis  meta-data (e.g., object attributes such as known or unknown face, confidence, date and time, etc.) are sent to the Cloud for storage and further retrieval~(Circled~\circled{7} in Figure~\ref{fig:cloud}).

\fakeparagraph{Cloud Layer} Figure~\ref{fig:cloud} presents a logical flow of the Cloud layer and its interactions with the Edge layer. The functions at this layer are:

\fakeparagraph{Real-time Push Notification} It sends a real-time alert notification to the user when a motion is detected in the proximity of the doorbell. This function is implemented using three components: AWS Kinesis, AWS Lambda, and AWS SNS (Circled~\circled{8}--\circled{9} in Figure~\ref{fig:cloud}). An ingested event by AWS Kinesis triggers the AWS lambda function that sends a push notification to Amazon SNS.

\fakeparagraph{Persistent data storage and access} 
It receives image analytics meta-data from the Edge and provides a scalable storage to access data anywhere and anytime~(Circled~\circled{10} in Figure~\ref{fig:cloud}). Both Amazon S3 APIs and DynamoDB APIs provide simple interfaces to store and retrieve data using the Amazon online storage infrastructure. We use these APIs that are exposed by Amazon API Gateway Service~(Circled~\circled{11} in Figure~\ref{fig:cloud}), which acts as a single point of entry and accommodates direct requests from the mobile app.

\fakeparagraph{Conversational User Interface}
The primary reason of integrating this feature is to use the natural language to interact with the smart doorbell. We implement it using Alexa Voice Service~(Circled~\circled{12} in Figure~\ref{fig:cloud}), which can be triggered using various voice commands. Our custom Alexa skill triggers a set of lambda functions to query image analytics meta-data storage. Once the query result is computed in the Cloud, the results are sent back to the user through Alexa voice service.



\section{Demonstration}


We plan to demonstrate the following use cases of our Smart doorbell:

\fakeparagraph{Use case 1: Cloud-based object detection and recognition}  It demonstrates the live object detections by the doorbell. The system is initially at rest. An object entering the proximity of the doorbell enables the smart doorbell to start. This activity automatically triggers the object detection and recognition. We implement a mobile app dashboard that provides the detailed activities at the doorbell. The notification messages include face recognition~(including known and unknown persons) and object detection (e.g., noteworthy car, animal, etc.).


\fakeparagraph{Use case 2: Real-time push notifications} It demonstrates the ability of sending real-time alerts to the user when a motion is detected in the proximity of the doorbell.  We implement an interface for real-time push notification. The user receives alerts on his mobile application when a visitor is detected at the door. The user can respond to the notification or just "ignore" it. In addition to this, we implement the the video library. This interface of the mobile app lets the users review activities and events at the door at a later time in case the user misses the real-time alert.



\fakeparagraph{Use case 3: On-demand live streaming}  It demonstrates the ability of our smart doorbell system to see what is happening at the doorbell on demand, anytime and anywhere.    This mobile app implements a software component that can allow users a high-resolution video stream of the events at the front door using the HLS protocol. We implement an on-demand live video streaming that allows users to see what is happening at the doorbell.


\fakeparagraph{Use case 4:  Conversational User Interface}  It demonstrates our approach that involves a voice assistant functionality, which allows the users to ask questions about the smart environment that they are in.  The voice assistant system leverages the logged image analytics results to provide a meaningful response. We have implemented an Alexa skill that can be triggered using various voice commands (such as ``\textit{Alexa, tell me what is happening at the door?}'', ``\textit{Alexa, send me a snapshot of all the activities at my door today}'').

\begin{acks}
This publication has emanated from research supported by grants from the European Union’s Horizon 2020 research and innovation programme under grant agreement number 847577 (SMART 4.0 Marie Sklodowska-Curie actions COFUND) and from Science Foundation Ireland (SFI) under grant number SFI/16/RC/3918 (Confirm) cofunded by the European Regional Development Fund.
\end{acks}

\bibliographystyle{ACM-Reference-Format}
\bibliography{sample-base}


\begin{thebibliography}{9}


\ifx \showCODEN    \undefined \def \showCODEN     #1{\unskip}     \fi
\ifx \showDOI      \undefined \def \showDOI       #1{#1}\fi
\ifx \showISBNx    \undefined \def \showISBNx     #1{\unskip}     \fi
\ifx \showISBNxiii \undefined \def \showISBNxiii  #1{\unskip}     \fi
\ifx \showISSN     \undefined \def \showISSN      #1{\unskip}     \fi
\ifx \showLCCN     \undefined \def \showLCCN      #1{\unskip}     \fi
\ifx \shownote     \undefined \def \shownote      #1{#1}          \fi
\ifx \showarticletitle \undefined \def \showarticletitle #1{#1}   \fi
\ifx \showURL      \undefined \def \showURL       {\relax}        \fi
\providecommand\bibfield[2]{#2}
\providecommand\bibinfo[2]{#2}
\providecommand\natexlab[1]{#1}
\providecommand\showeprint[2][]{arXiv:#2}

\bibitem[\protect\citeauthoryear{Chauhan, Patel, Delicato, and
  Chaudhary}{Chauhan et~al\mbox{.}}{2016}]%
        {chauhan2016development}
\bibfield{author}{\bibinfo{person}{Saurabh Chauhan}, \bibinfo{person}{Pankesh
  Patel}, \bibinfo{person}{Fl{\'a}via~C Delicato}, {and}
  \bibinfo{person}{Sanjay Chaudhary}.} \bibinfo{year}{2016}\natexlab{}.
\newblock \showarticletitle{A development framework for programming
  cyber-physical systems}. In \bibinfo{booktitle}{\emph{2016 IEEE/ACM 2nd
  International Workshop on Software Engineering for Smart Cyber-Physical
  Systems (SEsCPS)}}. IEEE, \bibinfo{pages}{47--53}.
\newblock


\bibitem[\protect\citeauthoryear{{Chauhan}, {Patel}, {Sureka}, {Delicato}, and
  {Chaudhary}}{{Chauhan} et~al\mbox{.}}{2016}]%
        {7460669}
\bibfield{author}{\bibinfo{person}{S. {Chauhan}}, \bibinfo{person}{P. {Patel}},
  \bibinfo{person}{A. {Sureka}}, \bibinfo{person}{F.~C. {Delicato}}, {and}
  \bibinfo{person}{S. {Chaudhary}}.} \bibinfo{year}{2016}\natexlab{}.
\newblock \showarticletitle{Demonstration Abstract: IoTSuite - A Framework to
  Design, Implement, and Deploy IoT Applications}. In
  \bibinfo{booktitle}{\emph{2016 15th ACM/IEEE International Conference on
  Information Processing in Sensor Networks (IPSN)}}. \bibinfo{pages}{1--2}.
\newblock


\bibitem[\protect\citeauthoryear{CNBC}{CNBC}{2020}]%
        {CNBC2020}
\bibfield{author}{\bibinfo{person}{CNBC}.} \bibinfo{year}{2020}\natexlab{}.
\newblock \bibinfo{title}{How Amazon Is Trying To Stop Package Theft}.
\newblock \bibinfo{howpublished}{YouTube:
  https://www.youtube.com/watch?v=6TqBJyd4aHg}.
\newblock


\bibitem[\protect\citeauthoryear{Delaney}{Delaney}{2020}]%
        {Delaney2020}
\bibfield{author}{\bibinfo{person}{John~R. Delaney}.}
  \bibinfo{year}{2020}\natexlab{}.
\newblock \bibinfo{title}{The Best Video Doorbells for 2020}.
\newblock \bibinfo{howpublished}{PC Magazine Article,
  https://in.pcmag.com/home-security/118816/the-best-video-doorbells-for-2020}.
\newblock


\bibitem[\protect\citeauthoryear{Gyrard, Serrano, and Patel}{Gyrard
  et~al\mbox{.}}{2017}]%
        {GYRARD2017305}
\bibfield{author}{\bibinfo{person}{Amelie Gyrard}, \bibinfo{person}{Martin
  Serrano}, {and} \bibinfo{person}{Pankesh Patel}.}
  \bibinfo{year}{2017}\natexlab{}.
\newblock \showarticletitle{Chapter 11 - Building Interoperable and
  Cross-Domain Semantic Web of Things Applications}.
\newblock In \bibinfo{booktitle}{\emph{Managing the Web of Things}},
  \bibfield{editor}{\bibinfo{person}{Quan~Z. Sheng}, \bibinfo{person}{Yongrui
  Qin}, \bibinfo{person}{Lina Yao}, {and} \bibinfo{person}{Boualem Benatallah}}
  (Eds.). \bibinfo{publisher}{Morgan Kaufmann}, \bibinfo{address}{Boston},
  \bibinfo{pages}{305 -- 324}.
\newblock
\showISBNx{978-0-12-809764-9}
\urldef\tempurl%
\url{https://doi.org/10.1016/B978-0-12-809764-9.00014-7}
\showDOI{\tempurl}


\bibitem[\protect\citeauthoryear{Intizar, Patel, Kanti~Datta, and
  Gyrard}{Intizar et~al\mbox{.}}{2017}]%
        {intizar:emse-01644333}
\bibfield{author}{\bibinfo{person}{Muhammad Intizar}, \bibinfo{person}{Pankesh
  Patel}, \bibinfo{person}{Soumiya Kanti~Datta}, {and} \bibinfo{person}{Amelie
  Gyrard}.} \bibinfo{year}{2017}\natexlab{}.
\newblock \showarticletitle{{Multi-Layer Cross Domain Reasoning over
  Distributed Autonomous IoT Applications}}.
\newblock \bibinfo{journal}{\emph{{Open Journal of Internet of Things}}}
  \bibinfo{volume}{3} (\bibinfo{year}{2017}).
\newblock
\urldef\tempurl%
\url{https://hal-emse.ccsd.cnrs.fr/emse-01644333}
\showURL{%
\tempurl}


\bibitem[\protect\citeauthoryear{Khochare, Aravindhan, and Simmhan}{Khochare
  et~al\mbox{.}}{2019}]%
        {scalable-framework}
\bibfield{author}{\bibinfo{person}{Aakash Khochare}, \bibinfo{person}{K.
  Aravindhan}, {and} \bibinfo{person}{Yogesh Simmhan}.}
  \bibinfo{year}{2019}\natexlab{}.
\newblock \showarticletitle{A Scalable Framework for Distributed Object
  Tracking across a Many-camera Network}.
\newblock \bibinfo{journal}{\emph{CoRR}}  \bibinfo{volume}{abs/1902.05577}
  (\bibinfo{year}{2019}).
\newblock
\showeprint[arxiv]{1902.05577}
\urldef\tempurl%
\url{http://arxiv.org/abs/1902.05577}
\showURL{%
\tempurl}


\bibitem[\protect\citeauthoryear{Patel, Gyrard, Datta, and Ali}{Patel
  et~al\mbox{.}}{2017}]%
        {10.1145/3041021.3054736}
\bibfield{author}{\bibinfo{person}{Pankesh Patel}, \bibinfo{person}{Amelie
  Gyrard}, \bibinfo{person}{Soumya~Kanti Datta}, {and}
  \bibinfo{person}{Muhammad~Intizar Ali}.} \bibinfo{year}{2017}\natexlab{}.
\newblock \showarticletitle{SWoTSuite: A Toolkit for Prototyping End-to-End
  Semantic Web of Things Applications}. In
  \bibinfo{booktitle}{\emph{Proceedings of the 26th International Conference on
  World Wide Web Companion}} (Perth, Australia) \emph{(\bibinfo{series}{WWW '17
  Companion})}. \bibinfo{publisher}{International World Wide Web Conferences
  Steering Committee}, \bibinfo{address}{Republic and Canton of Geneva, CHE},
  \bibinfo{pages}{263–267}.
\newblock
\showISBNx{9781450349147}
\urldef\tempurl%
\url{https://doi.org/10.1145/3041021.3054736}
\showDOI{\tempurl}


\bibitem[\protect\citeauthoryear{Pathak, Patel, Kanani, Arya, Patel, and
  Ali}{Pathak et~al\mbox{.}}{2020}]%
        {iot-conf-2020}
\bibfield{author}{\bibinfo{person}{Tapan Pathak}, \bibinfo{person}{Vatsal
  Patel}, \bibinfo{person}{Sarth Kanani}, \bibinfo{person}{Shailesh Arya},
  \bibinfo{person}{Pankesh Patel}, {and} \bibinfo{person}{Muhammad~Intizar
  Ali}.} \bibinfo{year}{2020}\natexlab{}.
\newblock \showarticletitle{A Distributed Framework to Orchestrate Video
  Analytics Across Edge and Cloud: A Use Case of Smart Doorbell~(To be
  appeared)}. In \bibinfo{booktitle}{\emph{Proceedings of the 10th
  International Conference on the Internet of Things}} (Malmo, Sweden)
  \emph{(\bibinfo{series}{IoT 2020})}. \bibinfo{publisher}{ACM},
  \bibinfo{address}{New York, NY, USA}, Article \bibinfo{articleno}{1},
  \bibinfo{numpages}{8}~pages.
\newblock


\end{thebibliography}










\end{document}